\documentclass[5p]{elsarticle}

\usepackage{lineno,hyperref}
\usepackage{amsmath}
\usepackage{siunitx}
\usepackage[para]{threeparttable}
\usepackage{graphicx}

\modulolinenumbers[5]

\journal{Elsevier}

\begin{document}

\begin{frontmatter}

\title{The search for IR excess in low signal to noise sources} 

\author[ucla,csun]{Jonathon K. Zink}
\ead{jzink@astro.ucla.edu}

 \author[csun]{Damian J. Christian} 

\address[ucla]{Department Astronomy, University of California, Los Angeles, CA 90095-1547}
\address[csun]{Department of Physics and Astronomy, Califorina State University, Northridge, CA 91330-8268}

\begin{abstract}
We present sources selected from their Wide-field Infrared Survey Explorer (WISE) colors that merit future observations to image for disks and possible exoplanet companions. Introducing a weighted detection method, we eliminated the enormous number of specious excess seen in low signal to noise objects by requiring greater excess for fainter stars. This is achieved by sorting through the 747 million sources of the ALLWISE database. In examining these dim stars, it can be shown that a non-Gaussian distribution best describes the spread around the main-sequence polynomial fit function. Using a gamma Probability Density Function (PDF), we can best mimic the main sequence distribution and exclude natural fluctuations in IR excess. With this new methodology we re-discover 25 IR excesses and present 14 new candidates. One source (J053010.20-010140.9), suggests a 8.40 $\pm$ 0.73 AU disk, a likely candidate for possible direct imagining of planets that are likely fully formed. Although all of these sources are well within the current flux ratio limit of $\sim$10$^{-6}$ (Wyatt 2008), J223423.85+403515.8 shows the highest bolometric flux ratio ($f_d$=0.0694) between disk and host star, providing a very good candidate for direct imaging of the circumstellar disk itself. In re-examining the Kepler candidate catalog (original study preformed by Kennedy and Wyatt 2012), we found one new candidate that indicates disk like characteristics (TYC 3143-322-1). 
\end{abstract}


\end{frontmatter}

\section{Introduction}

Circumstellar disks are created from the remnant material of stellar formation. Young protostellar disks provide a method for distinguishing the age of its stellar host and help models converge in determining the exact mechanisms of planet formation. Current models suggest most protoplanetary disks will photoevaporate within $\sim$ 1-5 Myr (Alexander et al 2006a, b; Owen et al. 2010). This infancy in which the star has just begun fusion, but not yet shed its disk, is the key time in which planet formation occurs. Finding stars within this narrow window of the stars lifetime, provides a further glimpse into the mysterious cause of planet formation. Additional clues to planet formation have resulted from the many planetary systems with large dust disks (Kalas et al. 2008; Lagrange et al. 2010; Marois et al. 2008; 2010). The presence of holes, gaps, and azimuthal symmetries in the dust distribution may also indicate the presence of undiscovered planets. Although many studies have not found strong correlation between the presence of circumstellar disks and planets, newer \textit{Herschel} observations have suggested there is a correlation (Marshall et al. 2014; Kennedy et al. 2014; 2015). For an alternate view see Moro-Martín et al. 2015.
\par There have been many studies attempting to quantify the occurrence of IR excesses and their inferred disks in FKG and M type stars. The occurrence of excess IR emission at longer wavelengths (70 $\mu$m), than those found by the mid IR régime of this study, have been found to be 10-15\% (Beichman et al. 2006; Trilling et al. 2008), compared to a much lower rate of $\sim$1\% for 24 $\mu$m emission (Lawler et al. 2009). Expanding these samples to stars known to host planets has found similar or even slightly lower rate for the occurrence of IR excesses (Bryden et al. 2009). More recently, the Wide-field Infrared Survey Explorer (WISE) provides information on millions of stars at 22 $\mu$m and Morales et al. (2012) found nine planet-bearing stars with warm dust emission; this gives an excess incidence for planet-bearing of only 1\% for Main Sequence stars.
\par Here we have undertaken a study to select stars that provide evidence of a disk from the ALLWISE catalog. This study differs from Patel et al. (2014), who searched for dust disks in the \textit{Hipparcos} catalog utilizing WISE data, Avenhaus et al. (2012), who detected circumstellar disks around M dwarfs, and the Theissen et al. (2014) study, which sought to examine population synthesis among disk harboring stars, by focusing on low SNR sources (<15) and further accounts for reddening effects seen by high magnitude signals in the WISE database. We also re-examine the current list of Kepler candidates for possible excess candidates (initial study was performed by Kennedy and Wyatt 2012, known as KW12 from here forth). In Section 2, we present the target selection criteria and the WISE photometry used. In Section 3 we present the IR excess results, and a table of their important parameters. In Section 4 we investigate some of the key features of the candidates, present Spectral Energy Distributions (SEDs) for noteworthy sources. Finally, in Section 5 concluding remarks are provided.

\section{Utilizing WISE and 2MASS}

This study uses of the ALLWISE (Wright et al. 2010) and 2MASS Catalogs (Skrutskie et al. 2006). From WISE the available photometric filters are as follows: 3.4, 4.6, 12, and 22 $\mu$m, corresponding to W1-4 (W1 thru W4) respectively. The 2MASS filters utilized are: 1.2, 1.7, and 2.2 $\mu$m, corresponding to 2MASS J, H, and K$_s$. The main focus of this study relies on WISE photometry as the mid-IR region provides the greatest signal for disk excess detection. WISE photometry also minimizes variability between filters by imaging all four filters simultaneously. To avoid false data reduction from over saturated images we impose saturation limits for WISE filters, in accordance with Patel (2005), at 4.5, 2.8, 3.8, and -0.4 mag for W1-4 respectively. In the test of the Kepler candidates and various other catalogs used in this study, we employ the IRSA database. Using a \SI{7}{\arcsecond} search radius (\SI{6.5}{\arcsecond} is the FWHM radius of the WISE photometry), the equatorial coordinates from each list were synchronized using the “One to One Match” feature of IRSA. Several of the sources did not generate a strong enough signal for WISE detection, or did not fall into the detectable range of the WISE survey, and thus were not included in this study. 

\subsection{Weighted Detection Method}

	In order to utilize the WISE data with low SNR, we investigated the instrumental reddening effect seen near the detection limits. Using stars from the Tycho-2 catalog (Hog et al. 2000), which present a B-V < 0.1 mag, we compared measurements from the ALLWISE database to those from the 2MASS catalog (Skrutskie et al. 2006). This study imposes the saturation limits of 3 mag for the K$_s$ band (as suggested by the 2MASS supplement)\footnote{Explanatory Supplement to the 2MASS All Sky Data Release, \href{url}{http://www.ipac.caltech.edu/2mass/releases/allsky/doc/explsup.html}} and remove sources with 2MASS SNR < 10. Since these blue stars have peaked at considerably smaller wavelengths, the true detection difference between the 2MASS K$_s$ filter and the WISE filters should be negligible. Figure 1 shows the differences for the W3 and W4 bands, demonstrating the great deviation from null as the magnitudes become fainter. It can be seen that beyond 8.8 mag for the W4 band scarce data were available. A notable drop in the standard deviation at this point, further indicates a peak in the sample. Because of this evidence, we established 8.8 mag as the detection limit for measurements in the W4 band. Similarly, the maximum for the W3 band was determined at 12 mag, as indicated in Figure 1.  When tested against W2 and W1 no apparent reddening effect was seen for my sample, thus we adopt the values of Cutri et al. (2012) for these filters. The limits suggested, greatly exceeded the magnitude of any tested stars in this study.

\begin{figure}
\resizebox{\hsize}{!}{\includegraphics{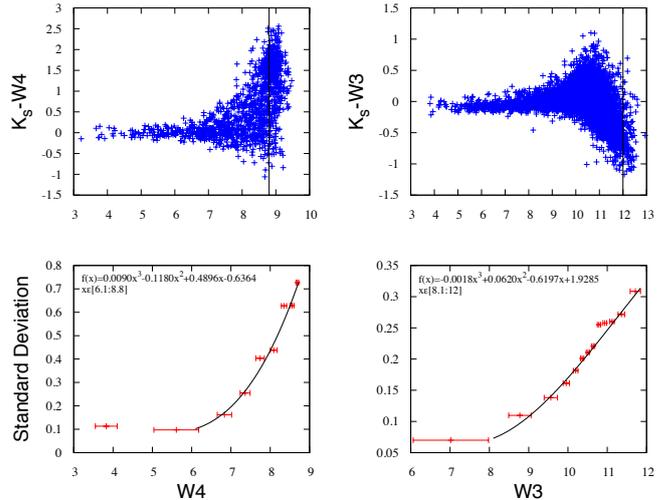}}
\caption{\textbf{Top} Two plots of Tycho-2 catalog stars with B-V $>$ 0.1 using the K$_s$ band from 2MASS and the W3, W4 bands of WISE (both in units of magnitude). All stars indicated a SNR $>>$ 5 in the K$_s$ band of 2MASS, while WISE bands varied down to SNR $>$2. 5-sigma clipping removed outliers to avoid contamination from excess sources and binary companions. The black line indicates the upper detection limits for respective filters within this study.
\newline \textbf{Bottom} Bins of 250 stars samples were taken from the upper plots and the standard deviation from the average was calculated. Deviation from the average was used since non-zero detection bias would already be accounted for by the main sequence polynomials (see Figure 2). The bars represent the standard deviation of data used for each sample. The function displayed represents that best fit line. I have adjusted these functions to reach zero at the lower bounds where normal photometric error becomes dominant. The lower bound was determined by the minimum point of the third order polynomial fit. Upper bounds represent a maximum deviation and the point when the sample size became too small for accurate representation.  }
\end{figure}
	
\par With interest in the low SNR data, we have incorporated the standard deviations attained from Figure 1 into the data set. When comparing our reddening values with the sigma values provided by the WISE pipeline, a \textit{Pearson Correlation Coefficient} of $\rho$ = 0.31 was found ($\rho$ is a ratio of covariance and variaces for two variables). This mild correlation indicates, unsurprisingly, that the WISE pipeline has already accounted for such effects. Since IR excess searches are often riddled with false positives in the low signal-to-noise regions, we have imposed extra weight to the sigma values of star with higher magnitudes. For magnitude values within the ranges indicated on Figure 1 the photometric uncertainties were adjusted accordingly:

$$
\sigma=\sqrt{\sigma^2_{Photo}+\sigma^2_{Red}}
$$

\noindent where $\sigma_{Red}$ corresponds to the value produced by the functions in Figure 1 and  $\sigma_{Photo}$ corresponds to the photometric uncertainty provide by the WISE pipeline. This weighting ensures a conservative approach when looking for significant excess at high magnitudes and low signal to noise, with hopes of eliminating possible dust contamination. The weighted technique acts as a minimum requirement for the detection of significant excess. 

\subsection{Main Sequence Fitting}

	To find IR-excess, we calibrated the WISE data to a Main Sequence sample of W1-W2, W2-W3, and W3-W4 colors. This helps account for instrumentation bias from various magnitude measurements made by WISE. Over 1,500 B2-M2 Main Sequence stars where drawn from the SIMBAD database with available WISE colors. Cross-correlation was achieved by using a \SI{6.5}{\arcsecond} search radius (\SI{6.5}{\arcsecond} is the FWHM of the WISE photometry). Later type M dwarfs have a SED turn over near the WISE and 2MASS filters and deviate from the normal colors of FGK types stars and thus have been excluded. Binaries and variable stars were removed to avoid contamination. A fourth order polynomial fit was used in order to best mimic the trends of each color diagram. To focus on the parameter space of this search, we have eliminated stars whose magnitude is <6. Furthermore, a 5-sigma clipping was applied to the data in order to avoid outliers from unaccounted for disks or red giants. We found reduced $\chi^2$ values <1, indicating the goodness of the fits. Figure 2 displays the fit model imposed for each of the colors. 

\begin{figure*}
\centering
\includegraphics[width=17cm]{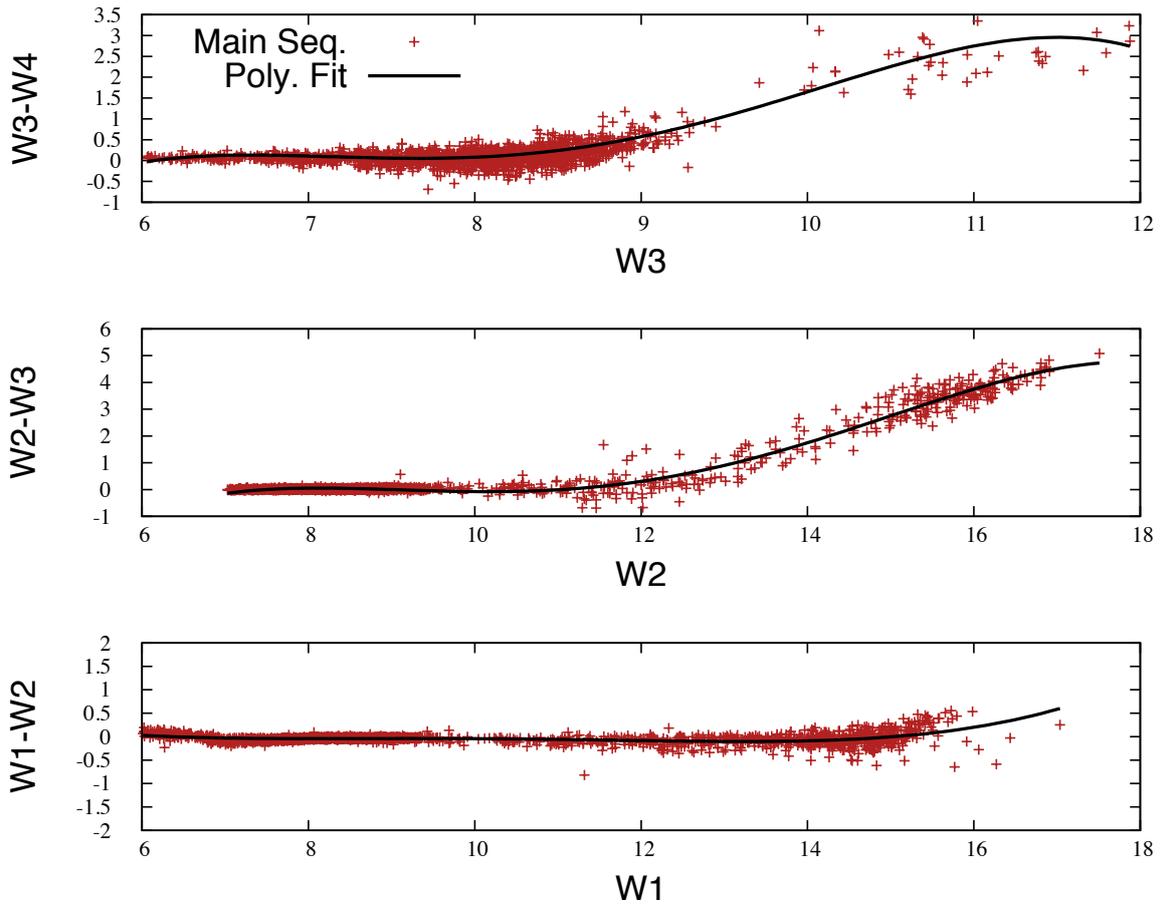}
\caption{Display of the three AllWISE color diagrams (all axis in units of magnitude). The fourth order polynomial main sequence fits are imposed over top of the data. This linear fit is due to detection response for faint objects and random galactic cirrus. These three diagrams provided the best fits ($\chi^2$< 1). As apparent, most stars lay along the trend line. Main sequence stars represent available data in SIMBAD and WISE catalog. Below 6 mag the stars follow a very narrow trend and force the fit defiantly, thus removal provides the best fit for high magnitude sources. }
\end{figure*}
	
\par Previous studies have noted the non-Gaussian features around Main Sequence fits (KW12). The skewed nature of these curves is likely due numerous instrumental and physical processes that produce increased IR flux. However, such non-uniformity has not yet been quantified. Many attempts have been made to use histograms to determine the parameters of these distributions, but these techniques rely on arbitrary bin widths imposed by the researcher. Here we have utilized the bin-less Quantile-Quantile plots (QQ plots) to better define the true nature of these distributions. Such plots utilize the inverse Cumulative Density Function (CDF) to linearize the data if the theorized CDF follows the true nature of the distribution. Performing an \textit{Anderson- Darling Normality Test} provides a measure of the probability the data comes from the tested distributions. Low probability (p) values indicates the need to rejected the tested model. We find p values <0.005 for all three color distributions, strongly advocating the need for an alternative model. In Figure 3, we demonstrate this need and show the strength of a gamma distribution for W2-W3 and W3-W4. The longer tails of the student t distribution provides a more robust model for the W1-W2 colors. Student t and gamma distributions are alternatives to the common Gaussian and provide unique features such as skew and elongated tails. Such components are necessary to fit non-normal data. While outliers still exist using these models, they have been minimized and are possibly due to non-typical main sequence phenomena. 
\par With the longer non-Gaussian tails suggested by the Gamma and T distributions, we can no longer assumed statistical significant for excess $>$3$\sigma'$ This study required 4.81$\sigma'$, 9.04$\sigma'$and 9.22$\sigma'$ for W3-4, W2-3, and W1-2 respectively, minimizing the chance of false positive. These values correlate to 4$\sigma'$, 4$\sigma'$, and 3$\sigma'$ levels of probability for a Gaussian distribution. In order to quantify the deviations we use a sigma test of:

$$
\sigma'=\dfrac{e}{\sigma}
$$

\noindent where $e$ is the deviation from the fit function and $\sigma$ corresponds to the weighted photometric error (for color measurements, error was added in quadrature to determine $\sigma$). For further discussion of weighted photometric error see section 2.1. This deviates from the normal form used in previous studies (Avenhaus, et al. 2012; Patel et al. 2014; Theissen et al. 2014), where the distributions were not well modeled and $\sigma'$ was a function of the $\chi^2$ from the fit polynomial. The strength of this method is the elimination of such dependence.

\begin{figure}
\resizebox{\hsize}{!}{\includegraphics{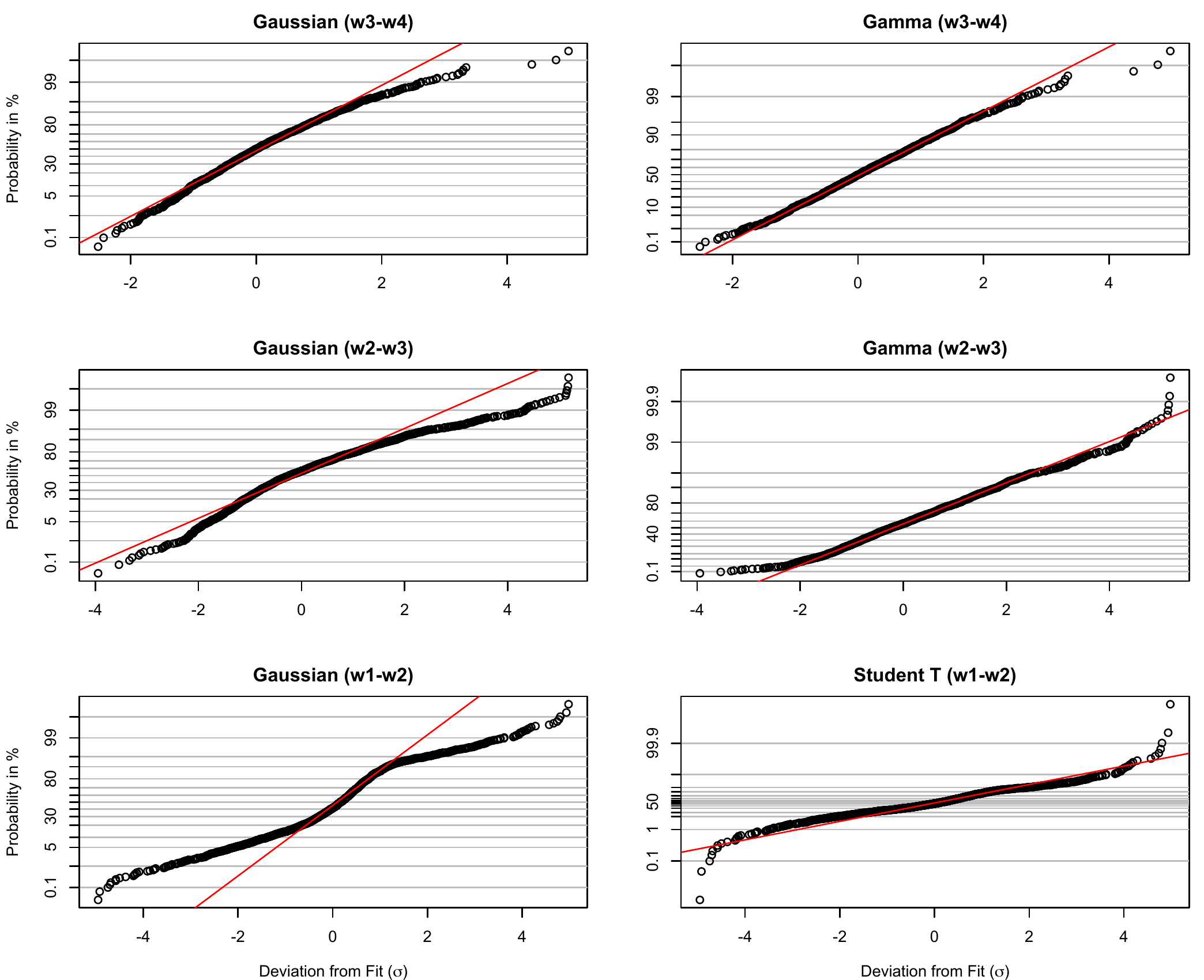}}
\caption{Probability plots for the Main Sequence distributions around the fit polynomials. The left graphs represent a Gaussian fit and the right graphs show the improved probability distribution fit. I found $\Gamma$($\alpha$=18, $\beta$=4.61), $\Gamma$(4, 1.47), and T(s=1.21, $df$=3) best fit the quantiles of the W3-4, W2-3, and W1-2 respectively. }
\end{figure}

\subsection{Catalog Requirements}

	The primary investigation incorporated the entire ALLWISE Catalog (747,634,026 sources). The initial cut in the catalog was made by removing signals with noted quality flags. The contamination and confusion flag (cc\_flag) indicates images affected by near bright stars or known artifacts. None zero indicators in the tested filters were removed. The extended source flag (ext\_flag) denotes the morphology quality and the source position fit within the 2MASS catalog. None zero values have been removed for the catalog. Beyond the $\chi^2$ <3 requirement of the ext\_flag we introduced a further constraint of  $\chi^2$ < 2, 1.5, 1.2, 1.2 for the respective W1-4 bands. This strict $\chi^2$ requirement minimizes the chances of contamination from possible binary signals or background galaxy (further discussion in section 2.4 and 2.5). Utilizing the variability flag (var\_flag) allowed us to account for variations from image to image for each filter. Indicators $>$ 5 have been removed from the tested filters. The moonlight flags (moon\_lev) helps account for moonlight contamination. Values $>$ 5 have been removed from the tested bands. Many of the flags are independent for each band. Thus, many sources may be removed from one test band while still being examined in others. 
\par Additional cuts have been made for stars without viable data from 2MASS. A SNR $>$ 2 was required for the H, J, and K$_s$ bands. Ideally, the ext\_flag would assure source correlation between WISE and 2MASS up to \SI{5}{\arcsecond}. To ensure this an additional distance check was made to remove any fraudulent matches.
\par Within the WISE catalog, photometric error values are provided for photometry with a SNR $>$2. Commonly, the images with 2<SNR<10 fall victim to background dust contamination and provide false IR excess signatures. Thus, previous studies sought to remove such contamination and often over look true excess in these regions of the sky. The goal was to provide evidence that such excesses are not due to dust contamination, but rather are true disk bearing stars. To maintain the low SNR examination goals of the WISE catalog a 2<SNR<15 requirement was imposed for this search. Although the region of 2$<$SNR$<$3 is extremely specious, statistical excesses could be found if an extraordinarily bright disk were to be detected in the parameter space. In total, we found 253, 43788, 34125 WISE sources met the stated requirements in W1-2, W2-3, W3-4 respectively.

\subsection{SED Fitting}

	To further ensure true stellar origins a simple two source blackbody model was initially fit to the candidate’s SED using WISE and 2MASS flux measurements. Models that produced fits with reduced $\chi^2$ $>$100 were removed. These are likely non-stellar objects such as galaxies or extragalactic flux. Late type M dwarfs have also been removed (T$_{eff}$ $<$ 3400K). Their low temperatures produced false positives when searching for FKG disks. 
\par Blackbody models are limited to only temperature parameters and provide little information on the type of star being observed. More sophisticated models include a surface gravity parameter ($\log{g}$) and metallicity (z), which helps eliminate red giants and young star forming regions from the catalog. Using the VOSA SED Analyzer (Bayo et al. 2008), we expanded the fits to include IRAS, SPITZER, AKARI, and SDDS photometry when available. Figure 4 displays an example of one such SED fit. Employing this expanded data set, we tested fits among AMES (Baraffe et al. 2003), Kurucz (Castelli et al. 1997), BT-Setti (Allard et al. 2012), BT-COND (Allard et al. 2009), and BT-NextGen (Allard et al. 2009) stellar models for best fit. we have provided parameters for the model that produced the smallest $\chi^2$ value. Utilizing these detailed models, we further removed sources with $\chi^2$ $>$10 (creating a finer filter from on our initial $\chi^2$ $>$100 cut).

\begin{figure}
\resizebox{\hsize}{!}{\includegraphics{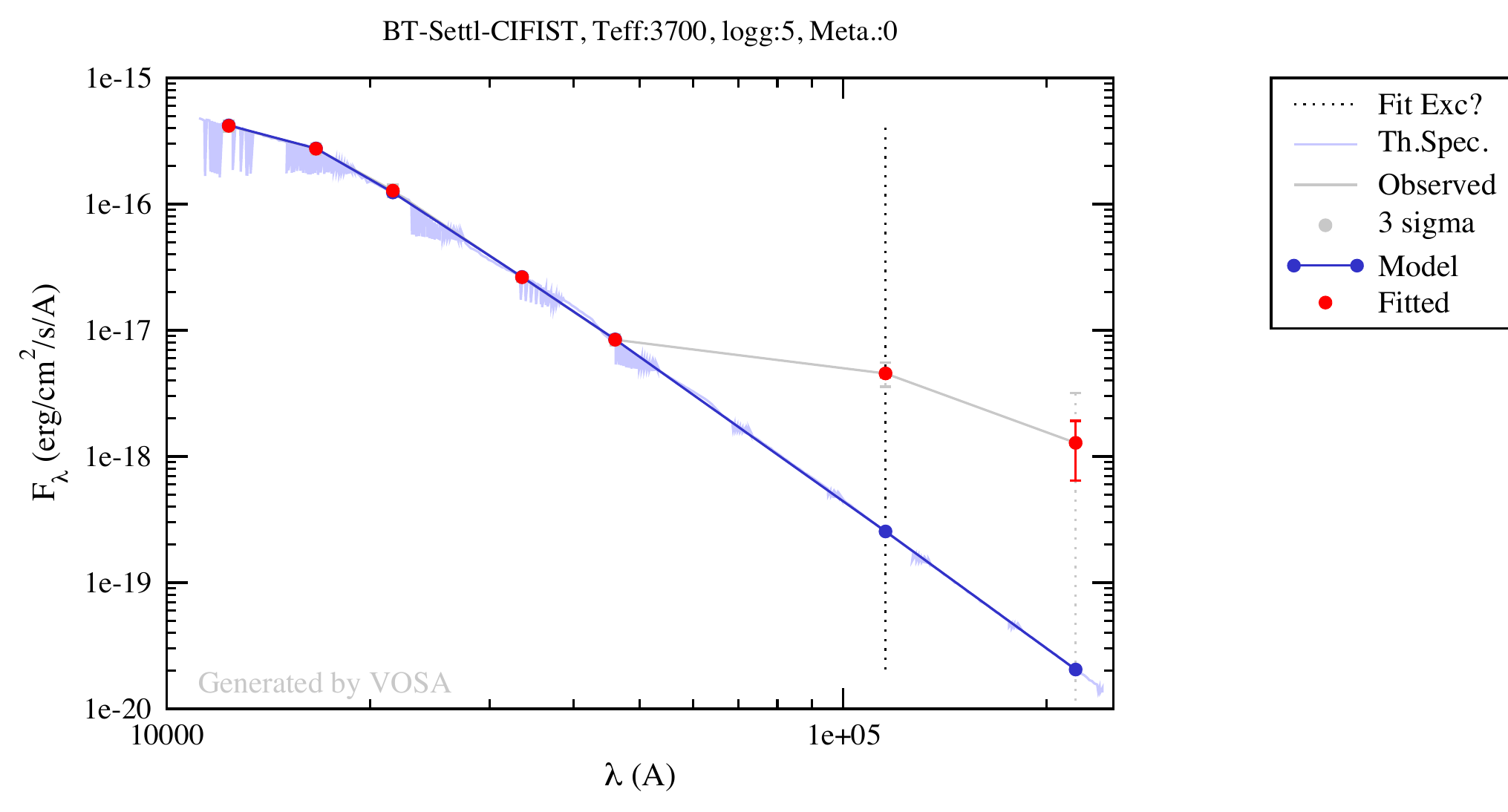}}
\caption{SED fit for J055138.64+024815.5, using the BT-Setti-CIFIST model. The light blue represents the actual model. The 7 signals are made from the three 2MASS bands and the 4 ALLWISE filters. Here the enormous excess produced in the W3 and W4 filter can be seen, thus qualifying this as an excess candidate.}
\end{figure}

\par With the possibility of binary contamination producing false IR excess, we have cross checked all sources with SIMBAD and removed any known binary systems. Furthermore, we have removed signals with possible contamination within \SI{15}{\arcsecond} of their listed coordinates. The NED database was employed to eliminate known galaxy contamination for a radius of \SI{15}{\arcsecond}.
\par Since IR excess search are often sensitive to red giants, eliminating such contamination is key for meaningful results. The SED models provide a $\log{g}$ value for the selected sources. Criteria for a non-giant (as suggested by Ciardi et al. 2011) are as follows: $\log{g}$ $>$ 3.5 for stars with T$_{eff}$ $>$ 6000K and $\log{g}$ $>$ 4 for stars with T$_{eff}$ $<$ 6000K. 
\par The nature of this study was to investigate low SNR regions, because of this requirement dust is of concern for all candidates. Galactic dust is known to contaminate the IR regime and produce false positives when searching for disk bearing stars. The W3-4 band is the most sensitive to dust contamination and, unfortunately, the most sensitive to IR excess. Previous studies, WK12 and Theissen et al. (2014) have utilized the 100 $\mu$m
filter of IRIS Data Collection Atlas to account for dust interference (Miville-Deschenes et al. 2005). Almost all of candidates presented here would fail the <5MJy requirement of these studies, even with Theissen’s recommendation to loosen the <5MJy parameter to $\sim$8MJy when studying sources away from the galactic disk. We have thus modeled the dust using all four filters of IRIS and a simple exponential fit. By doing so, we could interpolate the amount of dust contamination on each star and filter, providing a more precise measure of dust interference. Figure 5 shows the expected flux from dust contamination against the excess flux of the possible disks. Many of the W4 excess candidates live near the dust limit, indicating contamination rather than signal. Candidates were required to be at minimum 1$\sigma$ above y=x line in order to remove false positives due to dust reddening. 

\begin{figure}
\resizebox{\hsize}{!}{\includegraphics{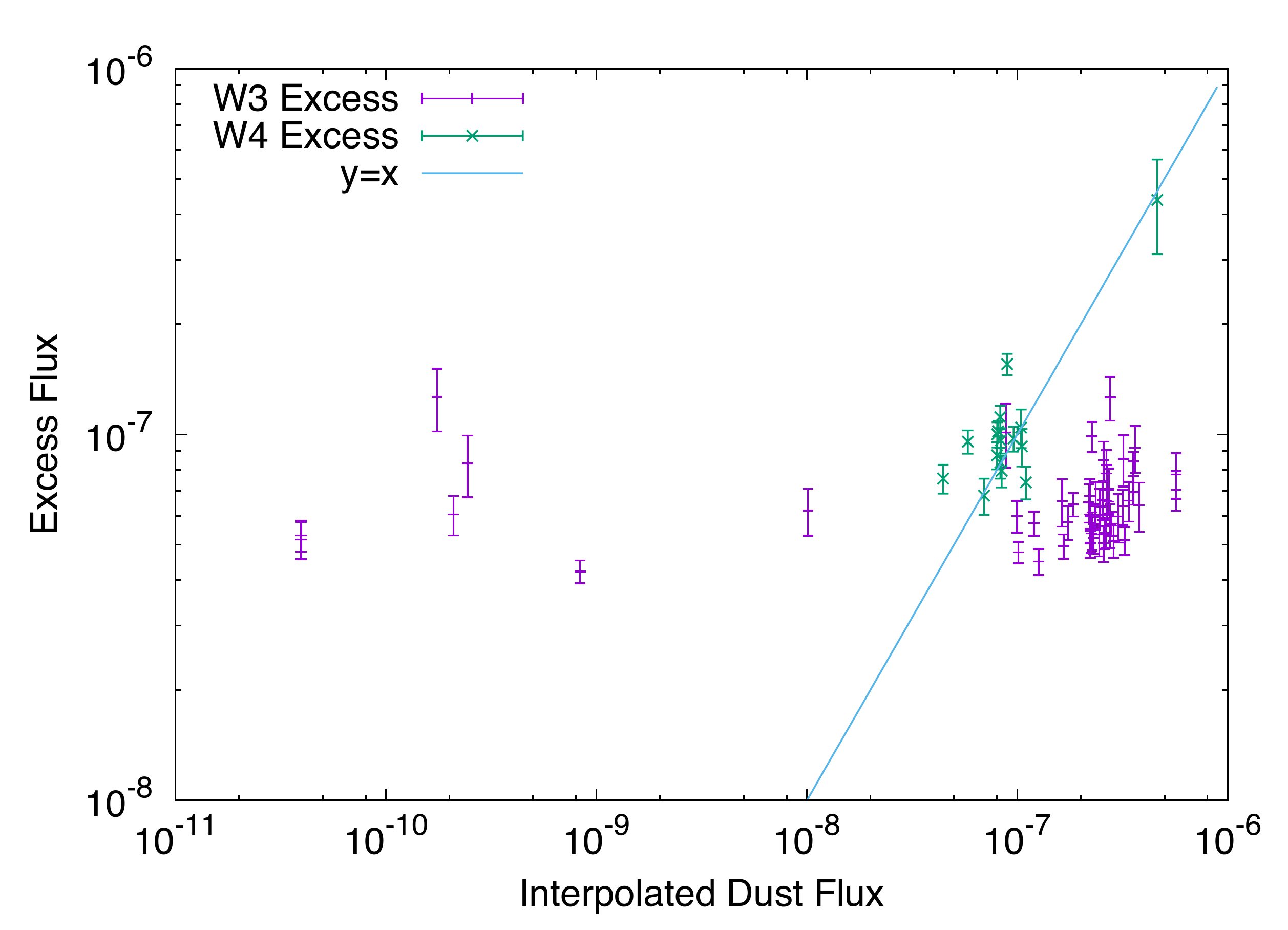}}
\caption{Comparison of best-fit dust contamination with excess flux. Flux measurements are in units of ergs/(s cm2). An exponential model was implemented to determine the flux at a the desired wavelengths. Many of the Interface Region Imaging Spectrograph (IRIS) 12$\mu$m data was omitted because of interference due to non-granule radiation.}
\end{figure}

\par Possible galactic alignment and extragalactic contaminants must also be considered. It was hoped that the small photometric $\chi^2$ constraints (discussed in Section 2.3) would eliminate any such alignment. However, we consider the possibility of alignment producing a point-source detection. Work done by Yan et al. (2013), suggests that mid to high galactic latitude ($| b |$ $>$\SI{20}{\degree}) regions could be contaminated by 1235 extragalactic sources per deg.2 in the W3 band. WISE photometry uses a \SI{6.5}{\arcsecond} FWHM, indicating $\sim$ 11 contaminates per object. Yan shows that this background provides a limiting magnitude value of $\sim$12.8 for W3. As specified in section 2.1, the maximum W3 magnitude was cut at 12 from our reddening calculation. Even more so, the largest W3 magnitude found in any of the candidates is $\sim$10.4, far above any suggested extragalactic background contamination. Any such pollution would have unobtrusive effects on the selected sources. Recently, Theissen et al. 2014 showed (with a Monte Carlo simulation), that meaningful galactic alignment contamination could only affect signals with W3 $>$ 12.7 mag. Again, well above the tested magnitudes of this study.  

\section{Results}

This study re-discovers 25 known disk-bearing star, all of which reside in the well-vetted HD catalog (McDonald et al. 2012). Beyond the previously known stars, 14 undiscovered excess candidates have been established (see Table 1). Interestingly, 5 objects are M-types disk candidates with SED temperatures ranging from 3500-3800K. A bias towards the discovery of M types stars is likely due to the conservative SNR restrictions. These low temperature disk candidates are rare and could provide idea candidates for imaging, due to the minimal star flux. Despite M dwarfs constituting a large majority of the stellar population ($\sim$70\% ; Bochanski et al. 2010), several unknown issues still exist about planet formation around these low mass stars. The possibility for exoplanet detection is also notables as the small size of these stars provides the necessary radius ratio for transit detections. Further discussion of disk parameters in section 4.2.

\subsection{Re-visiting Kepler}

By relaxing the SNR constrains to include all filters with SNR $>$ 2, we tested the current list of Kepler candidates (4,696 sources as of July 2015)\footnote{Kepler Candidates, \href{url}{http://www.nasa.gov/ames/kepler/kepler-planet-candidates-july-2015}}. As reported by KW12, we did not find any significant excess utilizing the requirements in section 2. Recent evidence suggests that the current Kepler pipeline (July 2014) is faulted and is discarding false negatives at rates greater than expected.\footnote{Kepler Global Erratum for Short Cadence, \href{url}{http://keplerscience.arc.nasa.gov/data/documentation/KSCI-19080-001.pdf}} Future adjustments to the pipeline may produce a large number of low SNR stars, which could be detected by the procedures listed. However, if we relax several of the candidate requirements we find one new star that warrants further study for disk candidacy (TYC 3143-322-1).

\begin{figure}
\resizebox{\hsize}{!}{\includegraphics{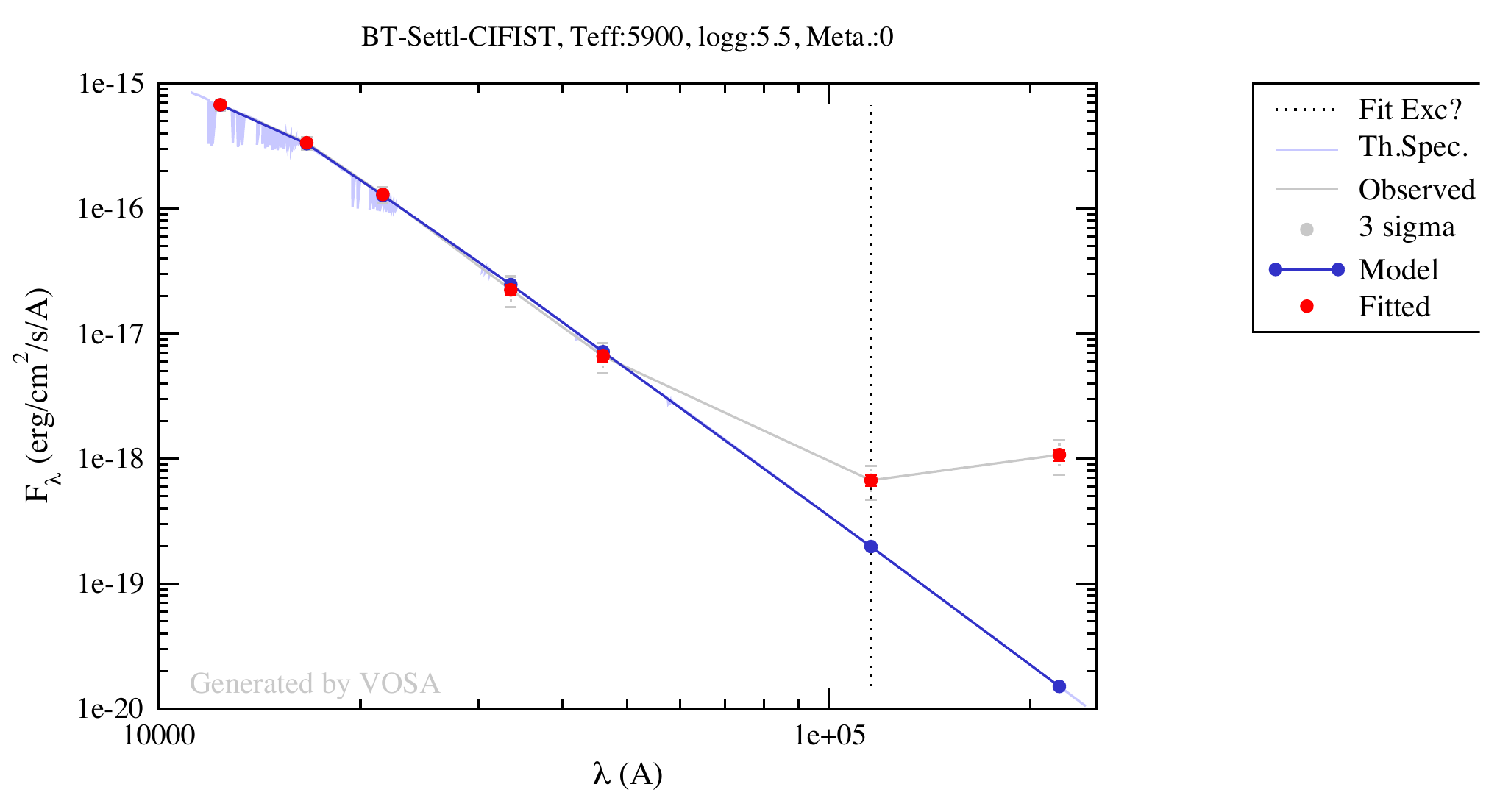}}
\caption{SED fit for TYC 3143-322-1, using the BT-Setti-CIFIST model. The light blue represents the actual model. The 7 signals are made from the three 2MASS bands and the 4 ALLWISE filters.}
\end{figure}

\begin {table*}
\caption {IR Excess Candidates}
\centering
    \begin{tabular}{ c  c  c  c  c  c  c  c }
\hline \hline
 & \multicolumn{1}{c}{RA$_{J2000}$} & \multicolumn{1}{c}{DEC$_{J2000}$}  &  \multicolumn{1}{c}{T$_{*}$} &   \multicolumn{1}{c}{T$_{disk}$}  &  \multicolumn{1}{c}{log g} &  \multicolumn{1}{c}{\emph{f}$_{d}$} &  \multicolumn{1}{c}{R$_{disk}$}\\
         \multicolumn{1}{c}{Identifier}  & \multicolumn{1}{c}{(deg.)} & \multicolumn{1}{c}{(deg.)} & \multicolumn{1}{c}{(K)} &  \multicolumn{1}{c}{(K)}  &  \multicolumn{1}{c}{ } & \multicolumn{1}{c}{(10$^{-3}$)} & \multicolumn{1}{c}{(AU)} \\

\hline  
\emph{W3 EXCESS} \\
J043600.30+062156.1$^{a}$     &  69.0012843 &  6.365605    &     4300$\pm$50         &      $<$230   &        5.5        &     32.0  & $>$0.55  \\
J223423.85+403515.8$^{b}$  &  338.5993826  &  40.5877494  &  4100$\pm$50    &     $<$240   &    5.5   &   69.4   &    $>$0.47  \\
J202232.25-252103.2$^{d}$   &   305.6343824   &   -25.3509126  &   3500$\pm$125    &        $<$230     &      5       &   42.7    &     $>$0.24  \\
J203111.66+100827.1$^{c}$    &  307.7986209   &    10.1408702   &   4500$\pm$50      &       230$\pm$75  &     5    &    39.2   &   0.59$\pm$0.39  \\
J052516.39-042207.2$^{b}$ & 81.3182992  &  -4.3686767  &   3500$\pm$50    &     $<$240     &    4.5   &     38.3   &  $>$0.22  \\
J192343.48-104736.0$^{c}$&  290.9311842  &   -10.7933553  &  5000$\pm$50   &        230$\pm$80    &     5.5   &      21.5   &   0.80$\pm$0.36  \\    
J181147.25-383330.3$^{b}$  &  272.9469063  &   -38.5584241  &  3800$\pm$50    &       $<$240    &     5    &      41.7  &     $>$0.39  \\  
J113115.72-201106.5$^{a}$  & 172.8155152  &  -20.1851401  &    5000$\pm$50   &       $<$230     &     5.5    &     35.2   &   $>$0.80  \\
\emph{W4 EXCESS} \\
J050442.72-041607.4$^{b}$   &     76.1780199  &   -4.2687383   &   4200$\pm$50   &     $<$270    &      5.5      &       29.1    &      $>$0.38\\
J205456.02+294513.5$^{c}$   &    313.7334307    &   29.7537656   &      4500$\pm$50   &  195$\pm$65  &      5.5   &   3.61     &  0.81$\pm$0.54\\
J191132.56-463926.1$^{b}$    &     287.8856889    &   -46.6572686   &    4200$\pm$50   &  $<$220  &      5.5    &     8.27    &    $>$0.58\\
J052707.76-012946.1$^{b}$   &     81.7823432   &     -1.4961575    &      3700$\pm$50   &  240$\pm$60    &       5.5   &    14.7    &   0.32$\pm$0.16\\ 
J053010.20-010140.9$^{a}$   &    82.5425082    &    -1.0280291    &      10000$\pm$100   & 230$\pm$10   &      5   &   0.637    &    8.41$\pm$0.73\\
J052330.49-044149.0$^{b}$   &    80.8770793    &     -4.6969545   &     3500$\pm$50   &  $<$255      &         4.5    &    49.1     &     $>$0.20\\
\hline
\end{tabular}
\flushleft
\begin{tablenotes}
\textbf{Fit Model:} $^{a}$ BT-NextGen, $^{b}$ BT-Settl-Cifist, $^{c}$ BT-Cond, $^{d}$ Kurucz
\end{tablenotes}
\end{table*}

\par We shall now list issues that have negated TYC 3143-322-1 from inclusion the KW12 survey. Sitting at a low galactic latitude ($| b |$$\sim$\SI{10}{\degree}), this signal lies in a region where 7.47 MJy/Sr of background was detected by the IRIS 100 $\mu$m map. Although not within the 5 MJy/Sr threshold of KW12, the enormous excess seen in the W3 and W4 band is unlikely due to this minor dust contamination. By interpolating the dust contamination from the 4 bands of IRIS, TYC 3143-322-1 was found to have an adjusted W4 value of 7.48 mag. This shows that the excess in the W4 band exists above the contamination level and is well within the detection limits of WISE. Figure 6 shows an SED of the apparent excess found for this source. The excessive $\chi^2$ values indicated in the W2, W3, W4 band are just above the KW12 limits and well within the limits of an expectable point-source for the WISE catalog ($\chi^2$ $<$3 in all bands). A galaxy alignment is possible, but unlikely to account for the excess in such a low magnitude star (W3$\sim$10 mag). Visual inspection further excludes such an alignment (see Figure 7). This star also has null values in terms of variability, likely due to only being imaged by WISE once. Future photometry may indicate no such variability. Because of these reasons, we believe this star should be considered a candidate for harboring a disk.  

\section{Discussion}

Detections of additional long-period extra-solar planet (ESP) systems are needed to constrain current planet formations theories and system architectures; dust disk detections have played a key role in these efforts (Smith and Terrile 1984; Kalas et al. 2008). Thus, sources with IR excesses may provide a short cut to targets with a good chance of also hosting long-period planets. Such deep imaging studies have been undertaken and are currently underway (Rameau et al. 2013; Wahhaj et al. 2015).  

\subsection{Disk Parameters}

Looking at the bolometric flux ratio of each star and excess, the detection limit for this study can be determined. 

$$
f_d=\dfrac{\int F_{IR} d\nu}{\int F_{\star} d\nu}
$$

\noindent where $f_d$ represents the bolometric flux ratio of the fit model for the star ($\star$) and the blackbody excess (IR).  $F_{IR}$ and $F_{\star}$ denote the flux values for the corresponding models. From the candidate list only 14\% of the stars presented $f_d$ values $<$ 0.005 with a minimum of $f_d$ = $6.37x10^{-4}$ for J053010.20-010140.9. A visual display of the WISE photometry for this source can be seen in Figure 7. This result is significantly lower than those reported by Patel et al. (2014), which is due to a more conservative adjustment to the photometric $\sigma$ values as discussed in Section 2. These dim excesses are defining properties of debris disks, which represent late disk evolution (Wyatt 2008; 2010). This indicates that a 87\% of theour candidates are young thick protoplanetary systems. With several unanswered questions in the field of planet formation, these candidates could potentially provide clues to construction in young dust disks.

\par One source, J223423.85+403515.8 provides this study’s highest $f_d$ value (see Table 1). With the high flux ratio, imaging should be routine and achievable. This early stage disk may provide a view into the early planet formation mechanisms.  
\par Assuming the disk acts as a perfect blackbody, an approximate disk radius can be calculated using:

$$
r=\left(\dfrac{278.3}{T_{IR}}\right)^2 \sqrt{L_{\star}}
$$

\noindent where $L_{\star}$ is in solar luminosity, $T_{IR}$ is the peak disk temperature in Kelvin, and $r$ is in astronomical units. Low temperature excess with high stellar luminosity, such as J053010.20-010140.9, provide ideal candidates for disk and possible exoplanet imaging. The low luminosity of the disk suggests that much of the disk has evaporated and planet formation is likely complete.  For this system we can approximate a disk radius $\sim$ 8.4 AU, just within the régime of current imaging capabilities (Maire et al. 2015). Table 1 also presents all of the calculated disk radius values. A vast majority of sources only provide a lower boundary for disk radius. With few data points in the mid-IR spectrum, the disk fits at very degenerate and may be more complex than a single black body fit as shown by Morales (et al. 2013). Direct imaging is required to verify a disk and it’s apparent radius. 
\par Although bright planets must exist in order for successful exoplanet imaging, $\beta$ Pic. (V $\sim$ 3.9) has provided famously effective imaging of its debris disk and exoplanet (Pepe et al. 2014).  Chances of imaging greatly improve for hotter planets (in younger systems) and those with significant angular separation. Utilizing the apparent relationship between large orbital distances and mid-IR excess, these candidates offer promising detection targets for new direct imaging observations. 

\section{Conclusion}

This study presents 14 candidates that show statically significant IR excess near the sensitive limits of WISE using this weighted method. These are likely due to dust disks surrounding host stars. Each candidate has been thoroughly vetted to ensure true excess over erroneous possibilities. Further combing of the Kepler candidates, found one star with disk like features that provide robust excess over background contamination. This star warrants further examination by direct imaging follow up. Of the systems discovered, one presents a large disk orbit 8.40 $\pm$ 0.73 AU and makes a good target for future deep imaging with AO and coronagraphs to search for new exoplanets. Several other stars provide lower bounds on disk radius and may harbor much larger disk, capable of imaging detection. This study provides evidence that true disk sources are abundant in these low SNR regions and merit continued study.

\begin{figure}
\resizebox{\hsize}{!}{\includegraphics{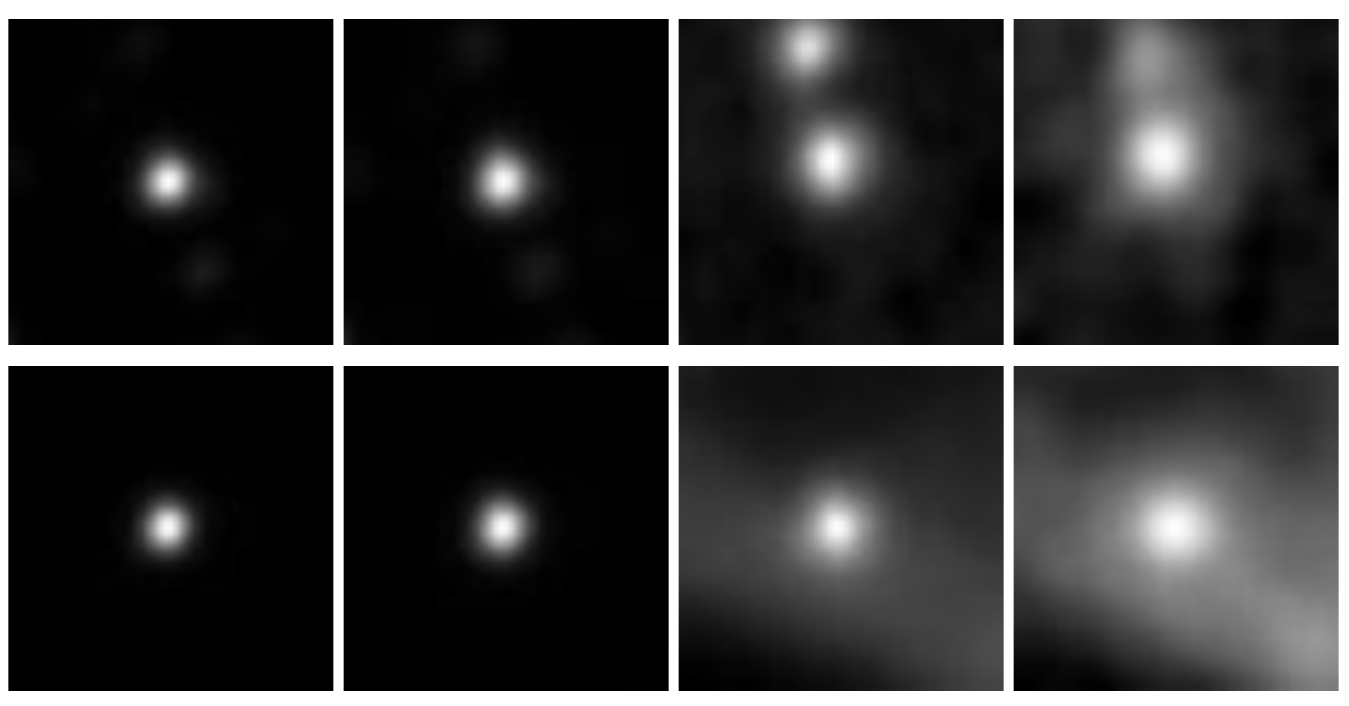}}
\caption{Four WISE waveband images for TYC 3143-322-1 (Top) and J053010.20-010140.9 (Bottom) . Represented from left to right are W1-4 bands respectively. These images show the mild contamination in these bands, indicating a true point source. A \SI{75}{\arcsecond} x \SI{75}{\arcsecond} portion of the sky is displayed in each image.  }
\end{figure}

\section*{Acknowledgements}
We would like to thank the CSUN Department of Physics and Astronomy for support of this project. A special thanks is also due to Farisa Morales for useful discussion on the WISE photometry and disk systems. This research has made use of the NASA/ IPAC Infrared Science Archive, which is operated by the Jet Propulsion Laboratory, California Institute of Technology, under contract with the National Aeronautics and Space Administration, the Exoplanet Orbit Database and the Exoplanet Data Explorer at exoplanets.org, and the Two Micron All Sky Survey, which is a joint project of the University of Massachusetts and the Infrared Processing and Analysis Center/California Institute of Technology, funded by the National Aeronautics and Space Administration and the National Science Foundation.

\section*{References}

\noindent \hangindent=.5cm
Alexander R. D., Clarck C. J., \& Pringle J. E. 2006a MNRAS, 369, 216

\noindent \hangindent=.5cm
Alexander R. D., Clarck C. J., \& Pringle J. E. 2006b MNRAS, 369, 229

\noindent \hangindent=.5cm
Allard, F., et al. 2001, ApJ 556, 357A

\noindent \hangindent=.5cm
Allard, F., et al. 2009, arXiv:1112.3591A

\noindent \hangindent=.5cm
Allard, F., et al. 2012, RSPTA 370. 2765A

\noindent \hangindent=.5cm
Avenhaus H., Schmid H. M., \& Meyer M. R. 2012 A\&A, 548, A105

\noindent \hangindent=.5cm
Baraffe, I., et al. 2003, A\&A, 402, 701B

\noindent \hangindent=.5cm
Baraffe, I., et al. 2015, A\&A, 577A, 42B

\noindent \hangindent=.5cm
Bayo, A., Rodrigo, C., Barrado y Navascués, D., Solano, E., Gutiérrez, R., Morales-Calderón, M., \& Allard, F. 2008, A\&A, 492, 277B.

\noindent \hangindent=.5cm
Beichman, C. A., Bryden, G., et al. 2006, ApJ, 652, 2

\noindent \hangindent=.5cm
Bryden, G., Beichman, C. A., et al. 2009, ApJ, 705, 2

\noindent \hangindent=.5cm
Castelli, F., et al. 1997, A\&A, 318, 841

\noindent \hangindent=.5cm
Ciardi, D. R., von Braun, K., \& Bryden, G., AJ, 141 108

\noindent \hangindent=.5cm
Cutri, R. M., Wright, E. L., Conrow, T., et al. 2012 Explanatory Supplement to the WISE All-Sky Data Release Products, Tech. Rep., NASA-IPAC

\noindent \hangindent=.5cm
Hog, E., Fabricius, C., Makarov, V. V., et al. 2000, A\&A, 355, L27

\noindent \hangindent=.5cm
Kalas, P., Graham, J. R., Chiang, E., et al. 2008, Science, 322, 5906 

\noindent \hangindent=.5cm
Kennedy, G. M., Wyatt, M. C., 2012 MNRAS, 426, 91

\noindent \hangindent=.5cm
Kennedy, G. M., et al. 2014, MNRAS, 438, L96

\noindent \hangindent=.5cm
Kennedy, G. M., et al. 2015, MNRAS, 449, 3121

\noindent \hangindent=.5cm
Lagrange, A.-M., Bonnefoy, M., Chauvin, G., et al. 2010, Science, 329, 57	

\noindent \hangindent=.5cm
Lawler, S. M., et al. 2009, ApJ, 705, 89

\noindent \hangindent=.5cm
Maire, A.-L., Skemer, A. J., Hinz, P. M., et al. 2015 A\&A, 576, A133

\noindent \hangindent=.5cm
Marois, C., Macintosh, B., Barman, T., Zuckerman, B., et al. 2008, Science, 322, 1348	

\noindent \hangindent=.5cm
Marois, C., Zuckerman, B., Konopacky, Q. M., et al. 2010, Nature, 468, 1080

\noindent \hangindent=.5cm
Marshall, J. P., et al. 2014, A\&A, 565, 15

\noindent \hangindent=.5cm
McDonald I., Zijlstra, A. A., \& Boyer, M. L. 2012, MNRAS, 427, 343

\noindent \hangindent=.5cm
Miville-Deschenes M., \& Lagache G. 2005, AJS, 157, 302

\noindent \hangindent=.5cm
Morales, F. Y., Bryden, G., Werner, M. W., et al. 2013, ApJ, 776, 111

\noindent \hangindent=.5cm
Morales, F. Y., Padgett, D. L., Bryden, G., Werner, M. W., \& Furlan, E. 2012, ApJ, 757,7

\noindent \hangindent=.5cm
Owen, J. E., Ercolano, B., Clarke, C. J., \& Alexander, R. D. 2010, MNRAS, 401, 1415 

\noindent \hangindent=.5cm
Pepe, F., Ehrenreich, D., \& Meyer, M. R. 2014, Nature, 513, 358

\noindent \hangindent=.5cm
Patel, R. J., Metchev, S. A, \& Heinze, A. 2014, ApJ, 212, 10

\noindent \hangindent=.5cm
Rameau, J., Chauvin, G., et al. 2013, ApJ, 772, L15

\noindent \hangindent=.5cm
Smith, B. A., \& Terrile, R. J. 1984, Science, 226, 1421

\noindent \hangindent=.5cm
Skrutskie, M. F., Cutri, R. M., Stiening, R., \& Weinberg, M. D.,  et al. 2006, AJ, 131, 1163

\noindent \hangindent=.5cm
Theissen, C. A., \& West, A. A., 2014, ApJ, 794, 146

\noindent \hangindent=.5cm
Trilling, D. E. et al. 2008, ApJ, 674, 1086 

\noindent \hangindent=.5cm
Wahhaj, Z., Liu, M.C., Nielsen, E. L., et al. 2015, ApJ, 773, 179

\noindent \hangindent=.5cm
Wright, E. L., Eisenhardt, P. R. M., Mainzer, A. K., et al. 2010, AJ, 140, 1868

\noindent \hangindent=.5cm
Wyatt, M. C., 2008, ARAA 46, 339   

\noindent \hangindent=.5cm
Wyatt, M. C., Booth, M., Payne, M. J., \& Churcher, L. J. 2010, MNRAS, 402, 657

\noindent \hangindent=.5cm
Yan, L., Donoso, E., Tsai, C. -W., et al. 2013, AJ, 145, 55

\end{document}